\documentclass[%
reprint,
superscriptaddress,
amsmath,
amssymb,
aps,
prx,
floatfix,
]{revtex4-2}
\usepackage{graphicx}% Include figure files
\usepackage{dcolumn}% Align table columns on decimal point
\usepackage{bm}% bold math
\usepackage{hyperref}% add hypertext capabilities
\usepackage{xr}
\usepackage{lipsum}
\usepackage{braket}
\usepackage[makeroom]{cancel}
\usepackage{xcolor}
\begin{document}
\preprint{APS/123-QED}
\bibliographystyle{unsrtnat} 
%=======================================================================================================================================
\title{A coherence sweet spot with enhanced dipolar coupling}
\author{J.\,H.~Ungerer}
\thanks{Equal contributions.}
\email{jungerer@g.harvard.edu}
\affiliation{
Swiss Nanoscience Institute, University of Basel, Klingelbergstrasse 82 CH-4056, Switzerland
}
\affiliation{
Department of Physics, University of Basel, Klingelbergstrasse 82 CH-4056, Switzerland
}
\affiliation{Department of Physics, Harvard University, Cambridge, MA 02138, USA}
\author{A.~Pally}
\thanks{Equal contributions.}
\affiliation{
Department of Physics, University of Basel, Klingelbergstrasse 82 CH-4056, Switzerland
}
\author{S.~Bosco}
\affiliation{
QuTech and Kavli Institute of Nanoscience, Delft University of Technology, Lorentzweg 1, 2628 CJ Delft, Netherlands
}
\author{A.~Kononov}
\affiliation{
Department of Physics, University of Basel, Klingelbergstrasse 82 CH-4056, Switzerland
}
\author{D.~Sarmah}
\affiliation{
Department of Physics, University of Basel, Klingelbergstrasse 82 CH-4056, Switzerland
}
\author{S.~Lehmann}
\affiliation{Solid State Physics and NanoLund, Lund University, Box 118, S-22100 Lund, Sweden}

\author{C.~Thelander}
\affiliation{Solid State Physics and NanoLund, Lund University, Box 118, S-22100 Lund, Sweden}
\author{V.F.~Maisi}
\affiliation{Solid State Physics and NanoLund, Lund University, Box 118, S-22100 Lund, Sweden}
\author{P.~Scarlino}
\affiliation{Institute of Physics and Center for Quantum Science and Engineering, Ecole Polytechnique Fédérale de Lausanne, CH-1015 Lausanne, Switzerland}
\author{D.~Loss}
\affiliation{
Department of Physics, University of Basel, Klingelbergstrasse 82 CH-4056, Switzerland
}
\author{A.~Baumgartner}
\affiliation{
Swiss Nanoscience Institute, University of Basel, Klingelbergstrasse 82 CH-4056, Switzerland
}
\affiliation{
Department of Physics, University of Basel, Klingelbergstrasse 82 CH-4056, Switzerland
}
\author{C.~Sch{\"o}nenberger}
\thanks{\url{nanoelectronics.unibas.ch}}
\affiliation{
Swiss Nanoscience Institute, University of Basel, Klingelbergstrasse 82 CH-4056, Switzerland
}
\affiliation{
Department of Physics, University of Basel, Klingelbergstrasse 82 CH-4056, Switzerland
}
%\date{\today}
\begin{abstract}
Qubits require a compromise between operation speed and coherence.
Here, we demonstrate a compromise-free singlet-triplet (ST) qubit, where the qubit couples maximally to the driving field while simultaneously coupling minimally to the dominant noise sources.
The qubit is implemented in a crystal-phase defined double-quantum dot in an InAs nanowire.
Using a superconducting resonator, we measure the spin-orbit interaction (SOI) gap, the spin-photon coupling strength and the qubit decoherence rate as a function of the in-plane magnetic-field orientation.
We demonstrate a spin qubit sweet spot maximizing the dipolar coupling and simultaneously minimizing the decoherence.
Our theoretical description postulates phonons as the most likely dominant noise source.
The compromise-free sweet spot originates from the SOI suggesting that it is not restricted to this material platform, but might find applications in any material with SOI.
These findings pave the way for enhanced engineering of these nanomaterials for next-generation qubit technologies.
\end{abstract}
\maketitle
Qubits based on single localized spins in semiconductor nanostructures represent cutting-edge platforms for quantum information processing, boasting extended coherence times and benefiting from established industrial fabrication techniques~\cite{hansonSpinsFewelectronQuantum2007}.
Scaling-up electron spin qubits is challenging because of limitations in all-electrical control and their small electric and magnetic dipole moments.
Complex structures such as microstrips or micromagnets are required to facilitate qubit manipulation~\cite{veldhorstAddressableQuantumDot2014,yonedaRobustMicromagnetDesign2015}.
However, the inherent spin-orbit interaction (SOI) in confined semiconductor hole systems~\cite{bulaev2005spin,bulaev2007electric,maurandCMOSSiliconSpin2016,watzingerGermaniumHoleSpin2018,froning2021ultrafast} and in electrons in nanowires (NWs)~\cite{nadj-pergeSpinOrbitQubit2010,schroer2011field,peterssonCircuitQuantumElectrodynamics2012,vanweperenSpinorbitInteractionInSb2015,han2023variable} offers an alternative coupling mechanism: SOI couples spin and charge degrees of freedom, enabling electric-dipole spin resonance~\cite{rashba2003orbital,golovach2006electric,nowack2007coherent,van2013fast} and spin-cavity coupling~\cite{nadj-pergeSpinOrbitQubit2010,peterssonCircuitQuantumElectrodynamics2012,kloeffel2013circuit,crippa2019gate,bosco2022fully,han2023variable,yuStrongCouplingPhoton2023,ungerer2024strong,de2023strong}.
\begin{figure}[tb]
    \centering
    \includegraphics[width=\linewidth]{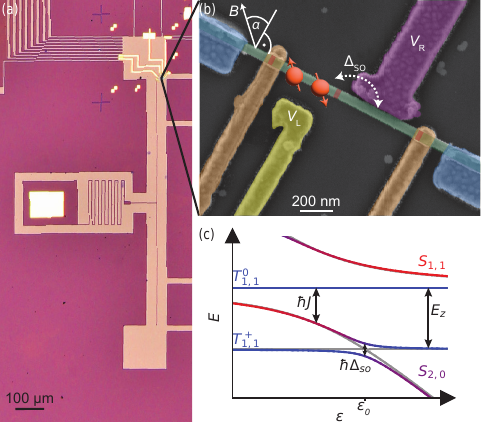}
    \caption{\textbf{Superconducting resonator coupled to singlet-triplet qubit in a crystal-phase NW.} \textbf{(a)} Optical microscope image of the device showing a half-wave NbTiN resonator with a characteristic impedance of 2.1\,k\textohm. At the middle of the center conductor, a dc bias line is connected via a meandered inductor. \textbf{(b)} False-colored scanning electron micrograph of the crystal-phase NW device (image is rotated by $-90^\circ$ with respect to (a)).
    The NW is placed at the position indicated in (a) and the purple gate line is galvanically connected to the resonator at its voltage anti-node. Tunnel barriers (red) are indicated with arrows.
    The device is operated with an even electron filling as depicted schematically.
    During the experiments, the in-plane magnetic field angle $\alpha$ is varied using a vector magnet.
    \textbf{(c)} Level diagram for an even electron occupation as a function of the electrostatic detuning $\varepsilon$ in the presence of strong SOI and finite magnetic field exhibiting singlet and triplet states.
    The avoided crossing between the spin-polarized triplet state $T^+$ and the low-energy singlet state at $\varepsilon=\varepsilon_0$ are detected using the resonator.}
    \label{fig:device}
\end{figure}

SOI is particularly relevant in singlet-triplet qubits, where it introduces a hybridization gap, for example between the singlet $|S\rangle$ and spin-polarized triplet $|T\textsuperscript{+}\rangle$ states~\cite{stepanenko2012singlet,nicholQuenchingDynamicNuclear2015,jirovec2021singlet}.
Recent experiments have demonstrated gigahertz-scale gaps, allowing to reach the strong coupling limit between these two-level systems and microwave resonators ~\cite{ungerer2024strong}.
These experiments extended the list of systems with strong spin-photon coupling~\cite{samkharadze2018strong,miCoherentSpinPhoton2018,landigCoherentSpinPhoton2018,yuStrongCouplingPhoton2023} to even charge parity states.

Despite several reports of SOI induced material characteristics as a function of the magnetic field orientation~\cite{piot2022single,yuStrongCouplingPhoton2023,han2023variable,d2013g,geyer2024anisotropic}, so far, only single sweet spots have been reported, either maximizing the coherence or the operation speed as a function of a global tuning parameter.
A recent report suggests that SOI could allow for simultaneous optimization of these two most relevant qubit parameters by local gates~\cite{carballido2024qubit}.
Here, we report that, in contrast to the common expectation, strong SOI allows one to simultaneously optimize both, the coherence and operation speed as a function of a global external parameter, here the angle of the applied magnetic field. We do so in a novel nanomaterial system with very reproducible characteristics.

We employ a magnetic-field resilient NbTiN resonator~\cite{samkharadzeHighKineticInductanceSuperconductingNanowire2016,ungerer2023performance} strongly coupled to a singlet-triplet (S-T\textsuperscript{+}) qubit in a crystal-phase defined InAs NW~\cite{lehmannGeneralApproachSharp2013,nilssonSingleelectronTransportInAs2016,ungerer2024strong}.
We apply an in-plane magnetic field in different orientations relative to the double quantum dot (DQD)~\cite{tanttu2019controlling,piot2022single,han2023variable,yuStrongCouplingPhoton2023}. The coherent coupling between the resonator and the singlet-triplet qubit allows to measure the qubit transition frequency, the vacuum Rabi rate, and the qubit decoherence rate.
Contrary to prevailing expectations, we identify a magnetic field orientation along the nanowire that serves as a compromise-free sweet spot, where a minimal decoherence rate coincides with a maximal dipolar coupling.
These findings are well explained by a theoretical model that suggests phonons as dominant source of decoherence.

This non-trivial optimization of two conflicting qubit parameters with only one external parameter is inherent to systems with a strong SOI and demonstrates the remarkable potential of such semiconductor materials.
Our results pave the way for future advances in material optimization and enhanced device functionality based on a deeper understanding of the underlying physics.
\section{Device}
The device is depicted in Fig.~\ref{fig:device}.
Detailed fabrication procedures are outlined in the Methods.
This device comprises a superconducting half-wavelength coplanar-waveguide resonator coupled to a DQD formed by in-situ crystal-phase engineering in an InAs NW ~\cite{nilssonSingleelectronTransportInAs2016,lehmannGeneralApproachSharp2013,ungerer2024strong,junger2020magnetic}.

Figure~\ref{fig:device}(a) shows an optical microscopy image of the resonator, fabricated by dry-etching an approximately 10\,nm thick NbTiN film, atop a SiO\textsubscript{2} substrate~\cite{ungerer2023performance}.
The small thickness of the superconducting NbTiN renders the resonator resilient to in-plane magnetic fields~\cite{samkharadzeHighKineticInductanceSuperconductingNanowire2016,ungerer2023performance}, and the narrow center conductor width, approximately 380 nm, combined with the large kinetic inductance of NbTiN, results in an impedance of 2.1\,k\textohm.
This large impedance enhances the vacuum electric field fluctuation amplitude compared to standard $50$\,\textohm-type resonators, thereby increasing the dipolar coupling to the qubit~\cite{stockklauser2017strong}.

The qubit is formed by electronic states in a DQD, with tunnel barriers grown deterministically by controlling the InAs crystal phase during the vertical growth process of the NW~\cite{lehmannGeneralApproachSharp2013,nilssonSingleelectronTransportInAs2016}.
The bariers are highlighted in the the colored SEM image of Fig.~\ref{fig:device}(b) in red and indicated by arrows.
The DQD forms within the zincblende segments of the NW, separated by wurtzite tunnel barriers.
At finite magnetic fields and in an even electron configuration, the DQD states are singlet and triplet states, as depicted in the energy level diagram in Fig.~\ref{fig:device}(c).
The ground state and first excited state comprise a superposition of the spin-polarized triplet $|T^+\rangle$, with an electron on each dot and the low-energy singlet state $|S_{2,0}\rangle$, with two excess electrons on one dot and none on the other.
As depicted in Fig.~\ref{fig:device}(c), without spin-flip tunneling, the energy levels would cross at a detuning $\varepsilon_0$ at which the Zeeman energy $E_z$ equals the exchange energy,
\begin{equation}
E_z=\hbar J\approx \frac{\hbar}{2}\left(\varepsilon_0+\sqrt{\varepsilon_0^2+4t_c^2}\right),
\label{eq:EzeqJ}
\end{equation}
where $t_c$ is the inter-dot tunnel rate.
But the finite inter-dot tunnel coupling and a substantial Rashba-type SOI in the zincblende segments~\cite{junger2020magnetic} result in a spin rotation and in the hybridization of the original eigenstates, with an energy gap of $\hbar\Delta_\mathrm{so}$~\cite{stepanenko2012singlet}.
The two hybridized levels constitute a spin qubit with a spin orbit-mediated electric dipole moment that couples to the electric field fluctuations of the resonator, rendering the resonator an effective probe for quantitative measurements of the qubit parameters.
\section{Hybrid system at large magnetic fields}
\begin{figure}[htb]
    \centering\includegraphics[width=\linewidth]{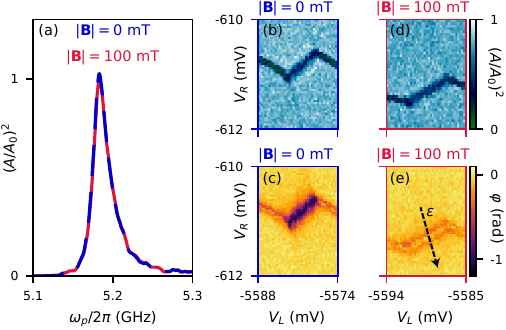}
    \caption{\textbf{Characterization of the hybrid device.} \textbf{(a)} Resonator transmission $(A/A_0)^2$ as function of probe frequency $\omega_p$. Fits to a lorenztian result in the resonance frequency $\omega_0/2\pi=5.1854\pm 0.0002$\,GHz, and decay rate $\kappa/2\pi=21.2\pm0.4$\,MHz independent of the magnetic-field amplitude $|\mathbf{B}|$. The magnetic-field resilience of the resonator enables resonator-based investigation of the DQD. \textbf{(b-e)} Transmission $(A/A_0)^2$ and phase $\varphi$ close to resonance frequency as a function of gate voltages $V_L$ and $V_R$ applied to the DQD as illustrated in Fig.~\ref{fig:device}(b) at zero field and at $|\mathbf{B}|=100$\,mT. Position and shape of the observed inter-dot transition signal varies as function of $|\mathbf{B}|$ due to an interaction of the resonator with spinful DQD levels.}
    \label{fig:honeycomb}
\end{figure}
In our experiments, we measure the amplitude $A$ and phase $\varphi$ of a microwave probe tone transmitted through the resonator, capacitively coupled to the DQD.
In Fig.~\ref{fig:honeycomb}(a), the squared normalized transmission amplitude $(A/A_0)^2$ through the resonator is plotted as a function of the probe frequency $\omega_p$ with the DQD tuned into Coulomb blockade, rendering it irrelevant for the measurement.
This figure displays probe frequency scans at $|\mathbf{B}|=0$\,mT and $|\mathbf{B}|=100$\,mT, where the field is applied in-plane.
As visible in Fig.~\ref{fig:honeycomb}(a), the transmission though the resonator recorded for $|\mathbf{B}|=0$\,mT and $|\mathbf{B}|=100$\,mT does not show any variation, demonstrating that the resonator remains unaffected for the magnetic fields used in our experiments.

We now prepare the DQD in an even charge state ~\cite{schroer2012radio,malinowski2018spin,ezzouch2021dispersively,ungerer2024strong} and measure the resonator transmission at a frequency close to resonance.
Figures ~\ref{fig:honeycomb}(b), ~\ref{fig:honeycomb}(c) ~\ref{fig:honeycomb}(d), and ~\ref{fig:honeycomb}(e) depict the transmission amplitude and phase as functions of the gate voltages $V_L$ and $V_R$ for in-plane fields of $|\mathbf{B}|=0$ and $|\mathbf{B}|=100$\,mT.

Due to the electric-dipolar coupling between the resonator and the DQD, the resonator transmission directly reveals the charge stability diagram of the DQD~\cite{freyDipoleCouplingDouble2012}.
At the inter-dot transition (IDT), the Zeeman energy and the exchange energy are degenerate (see Eq.~\eqref{eq:EzeqJ}) and the hybridized states form a qubit that couples to the resonator.
The IDT can be easily identified, signaled by lines with positive slopes in Fig.~\ref{fig:honeycomb}(e).

Both, the position of the IDT in the gate-versus-gate map and the resonator response near the IDT strongly depend on the external magnetic field strength.
This susceptibility to magnetic fields arises from the spin-dependent DQD transitions. 
In the following, we probe the resonator response as a function of electrostatic detuning $\varepsilon$ which is manipulated by varying the gate voltages $V_L$ and $V_R$ along the line indicated in Fig.~\ref{fig:honeycomb}(e).
\section{A double sweet spot}
\begin{figure}
\includegraphics[width=\linewidth]{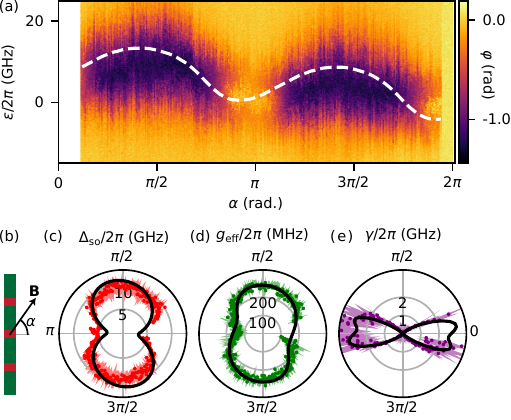}\caption{\textbf{Qubit parameters as function of in-plane field angle at $\mathbf{|B|=100}$\,mT.} \textbf{(a)} Phase of the resonator transmission $\varphi$ as function of electrostatic detuning $\varepsilon$ and in-plane magnetic field angle $\alpha$.
The detuning $\varepsilon$, was calculated from the applied voltages, using the gate-to-dot lever arms (see Tab.~\ref{tab:params}).
The dashed curve corresponds to the position of the IDT in the theoretical model, $\varepsilon=\varepsilon_0$, to which a linear trend was added accounting for a drift of the gate voltage.
\textbf{(b)} Schematic showing the alignment of the magnetic field $\mathbf B$ with respect to the nanowire. \textbf{(c)} SOI gap $\Delta_\mathrm{so}$ as function of $\alpha$. \textbf{(d)} Qubit-resonator coupling strength $g_\mathrm{eff}$ as function of $\alpha$.
\textbf{(e)} Qubit decoherence rate $\gamma$ as function of $\alpha$. 
When the field is parallel to the NW, $\alpha=\pm \pi/2$, a compromise-free sweet spot is found where maximal qubit transition frequnecy and coupling strength coincide with minimal decoherence. (c), (d) and (e) are extracted using input-output theory.
The streaks symbolize the uncertainty of the fit.
This uncertainty is a consequence of the uncertainty of the gate lever arms which forms the most significant source of uncertainty in our experiment and is stated in the caption of Tab.~\ref{tab:params}.
All curves overlaid on the data result from the theoretical model described in the main text, using a single set of fit parameters.}
    \label{fig:Brot100mT}
\end{figure}
The main result of our work is represented by the dependence of the IDT characteristics on the in-plane field orientation.
For this we show in Fig.~\ref{fig:Brot100mT}(a) the resonator transmission phase $\varphi$ close to the resonance frequency, plotted as a function of the in-plane angle $\alpha$ between the NW axis and the magnetic field, illustrated in Fig.~\ref{fig:device}(b) and Fig.~\ref{fig:Brot100mT}(b), and versus the electrostatic DQD detuning $\varepsilon$ as defined in Fig.~\ref{fig:honeycomb}(e).
Fig. \ref{fig:Brot100mT}(a) reveals that the detuning $\varepsilon_0$, at which the IDT is observed, varies as a function of $\alpha$.
This angle dependence can be understood qualitatively by recognizing that the SOI introduces an anisotropic g-tensor and hence determines the Zeeman energy $E_Z$ and, according to Eq.~\ref{eq:EzeqJ}, the position of the IDT $\varepsilon_0$.
Furthermore, the phase signal as a function of $\varepsilon$ changes continuously from a negative shift in $\varphi$ at $\alpha\approx \pi/2$~$[3\pi/2]$ to a double-dip structure at $\alpha\approx\pi$~$[2\pi]$, because the SOI renders the magnitude of $\Delta_\mathrm{SO}$ and the decoherence rate $\gamma$ angle-dependent by affecting the overlap of the spin wavefunctions~\cite{stepanenko2012singlet,tanttu2019controlling,piot2022single,hendrickx2023sweet}.
For example, at $\alpha\approx \pi/2$, $\Delta_\mathrm{so}>\omega_0$, resulting in a dispersive resonator signal.
In contrast, at $\alpha\approx\pi$, $\Delta_\mathrm{so}<\omega_0$ so that an (avoided) crossing between the singlet-triplet qubit and the resonator is observed as function of $\varepsilon$.
This crossing experimentally results in a double dip structure in the $\varphi(\varepsilon)$ dependence, framing a positive shift at the center of the IDT, at $\varepsilon\approx\varepsilon_0$.

Using input-output theory~\cite{gardiner1985input} as described in the Methods, we extract the SOI gap $\Delta_\mathrm{so}$, the qubit decoherence rate $\gamma$, as well as the effective spin-photon coupling strength $g_\mathrm{eff}$ from the dependence of the resonator phase and amplitude on $\varepsilon$.
The results are plotted as function of in-plane angle $\alpha$ in Fig.~\ref{fig:Brot100mT}(c),~\ref{fig:Brot100mT}(d) and~\ref{fig:Brot100mT}(e).

Excitingly, we find that, while $\Delta_\mathrm{so}$ and $g_\mathrm{eff}$ are correlated with each other, they are anticorrelated with $\gamma$.
In particular, when applying the magnetic field parallel to the NW, a compromise-free sweet spot is found, for which the spin-photon coupling strength $g_\mathrm{eff}$ is maximal while $\gamma$ is minimal.
To robustly estimate the spin-photon coupling and the decoherence rate at the sweet spot, we average over all extracted values in the interval $\alpha\in [55^\circ,125^\circ]\cup[235^\circ,305^\circ]$ and find $\langle g_\mathrm{eff}\rangle/2\pi=250\pm 15 (30)$\,MHz and $\langle\gamma\rangle/2\pi=11\pm 20 (400)$\,MHz, where the error corresponds to the statistical (systematic) uncertainty. Here, the large systematic uncertainty originates from the uncertainty of the gate lever arms stated in the caption of Tab.~\ref{tab:params}.
\section{Theoretical description}
In this section, we outline how the effective Hamiltonian yields the the dashed white and solid black curves shown in Figures~\ref{fig:Brot100mT}(a),~(c), and~(d). The consecutive section describes how the decoherence is modeled (black curve in Fig.~\ref{fig:Brot100mT}(e)). 

All curves are the result of numerically diagonalizing the Hamiltonian $H_5$ in Eq.~\eqref{eq:H5} in the Methods that describes the DQD states in proximity to the IDT.
Thereby, we take into account an anisotropic g-tensor, identical for both dots, and a spin-orbit vector that is not aligned with the principal axes of the g-tensor.
We point out that all curves are obtained from a single set of fit parameter, given in Table~\ref{tab:params}.
The Landé g-factor varies between $g=10.5$ and $g=5.25$, depending on the field direction. And, taking into account the distance of the two quantum dots, $d=330$\,nm, the spin-orbit length is found as $l_\mathrm{so}=130$\,nm.
These values are consistent with previous experiments~\cite{fasth2007direct,nadj-pergeSpinOrbitQubit2010}.
In Fig.~\ref{fig:g-tensor} in the Methods, we use the fitted parameters to visualize the g-tensor.

After finding the eigenenergies and eigenstates from diagonalizing the Hamiltonian $H_5$, we focus on the ground state $\ket{0}$ and first excited state $\ket{1}$ with their respective energies $E_0$ and $E_1$. These form the qubit states which are sensed by the resonator.
Because Eq.~\eqref{eq:EzeqJ} is only valid for small SOI, which is not the case for certain field alignments, we more generally determine the detuning $\varepsilon_0$, at which the anti-crossing occurs, by identifying the minimum gap for which
\begin{equation}
    \partial_\varepsilon (E_1-E_0)=0.
\end{equation}
The numerical solution for $\varepsilon_0$ is plotted in Fig.~\ref{fig:Brot100mT}(a) on top of the data (white dashed curve).
The variations of the detuning $\varepsilon_0$ of the IDT as function of $\alpha$ is caused by the g-tensor anisotropy.
According to Eq.~\eqref{eq:EzeqJ}, this results in a variation of $\varepsilon_0$ at which the Zeeman energy equals the exchange energy.
As a consequence, the detuning $\varepsilon_0$ at which the IDT is observed varies with the g-factor, which in turn is given by the field orientation.
Then, we numerically calculate the SOI gap \begin{equation}
    \Delta_\mathrm{so}=E_1(\varepsilon_0)-E_0(\varepsilon_0)\ ,
    \end{equation}
plotted as solid black curve in Fig.~\ref{fig:Brot100mT}(c).
The variation of the SOI gap $\Delta_\mathrm{so}$ is a consequence of the magnetic field orientation with respect to the anisotropic g-tensor and the spin-orbit vector.
To calculate the vacuum Rabi coupling strength $g_\mathrm{eff}$, we assume that the resonator couples to the electric dipole moment of the singlet-triplet qubit via the resonator vacuum fluctuations in the detuning of amplitude $\delta\varepsilon_0$ according to Eq.~\eqref{eq:coupling} in the Methods.
This gives rise to the dipolar coupling strength as the vacuum Rabi rate
\begin{equation}
g_\mathrm{eff}=\delta \varepsilon_0\left|\bra{0} h_{\delta\varepsilon}\ket{
1} \right|
\label{eq:geff}
\end{equation}
plotted as solid black curve in Fig.~\ref{fig:Brot100mT}(d). Here, $h_{\delta\varepsilon}$ is the operator describing small variations of the detuning, given by Eq.~\eqref{eq:delta_det} in the Methods and evaluated at the center of the IDT, $\varepsilon=\varepsilon_0$.
\section{Decoherence}
In the experiment, using input-output theory, we utilize the resonator as a probe to extract the qubit decoherence rate $\gamma$, which is plotted in Fig.~\ref{fig:Brot100mT}(e), and defined by
\begin{equation}
    \gamma=\frac{\gamma_1}{2}+\gamma_\varphi,
\end{equation}
where $\gamma_1$ is the relaxation rate and $\gamma_\varphi$ the pure dephasing rate.
The primary sources of decoherence in quantum dots are typically hyperfine-interaction induced dephasing from atomic nuclei~\cite{assali2011hyperfine,schliemann2003electron,testelin2009hole}, charge noise-induced dephasing~\cite{petersson2010quantum,holman20213d,paladino20141,scarlino2022situ}, or relaxation due to phonons~\cite{fujisawa1998spontaneous,golovach2004phonon,trif2008spin,kornich2014phonon,kloeffel2014acoustic,kornich2018phonon,hartke2018microwave,hofmann2020phonon,zou2024spatially}.

From our experiments, we extract an unexpected anti-correlation between the spin-photon coupling strength $g_\mathrm{eff}$ and the decoherence rate $\gamma$.
Using our theoretical description, and Bloch-Redfield theory~\cite{golovach2004phonon,cywinski2008enhance,kornich2018phonon}, we investigate various possible decoherence mechanisms as outlined in the Supplementary Material (SM).
We find that consideration of magnetic noise stemming from nuclear spins or charge-noise due to phonons leads to the correct trend of $\gamma$ as a function of the in-plane magnetic field angle $\alpha$.
However, because we identify decoherence rates comparable to the maximum of $\gamma(\alpha)$ also for a charge qubit at zero magnetic field, where magnetic noise is irrelevant, we hypothesize that phonons form the dominant noise source in our experiment. Therefore, here, we focus on phonon-mediated decoherence.

\begin{figure}
\includegraphics[width=\linewidth]{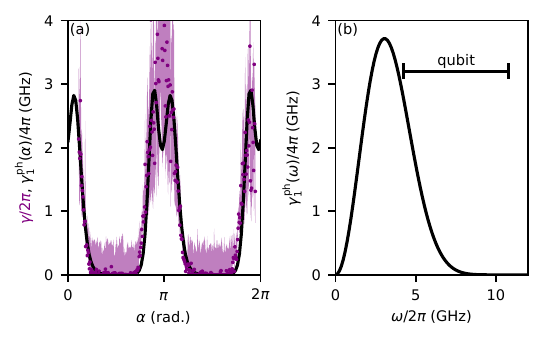}\caption{\textbf{Phonons as possible decoherence source.} \textbf{(a)} Measured decoherence rate $\gamma$ as a function of in-plane angle $\alpha$ (purple).
The black curve is the numerically calculated relaxation rate $\gamma_1^\mathrm{ph}(\alpha)$ originating from deformational phonons (see main text). \textbf{(b)} Analytical relaxation rate $\gamma_1^\mathrm{ph}(\omega)$ as function of phonon frequency $\omega$ (see SM for details).
The qubit operates in the frequency range as indicated with a negative slope of $\gamma_1^\mathrm{ph}(\omega)$. This explains the anti-correlation between the SOI gap $\Delta_\mathrm{so}$ and the decoherence rate $\gamma$ - a possible reason for the compromise-free sweet spot formation.}
\label{fig:gamma}
\end{figure}

Fig.~\ref{fig:Brot100mT}(e) and Fig.~\ref{fig:gamma}(a) show the phonon-mediated relaxation $\gamma_1^\mathrm{ph} (\alpha)$ as a function of the in-plane field angle $\alpha$ overlaid on the measured decoherence rate $\gamma$.
Considering only the effect of a single gapless, low-energy phonon band gives rise to the analytical functional dependence $\gamma_1^\mathrm{ph}(\omega)$ as function of phonon frequency, $\omega$ (see SM for the derivation).
This dependence is plotted in Fig.~\ref{fig:gamma}(b).
These phonons lead to qubit relaxation when their frequency is close to the qubit transition frequency, $\omega\approx\Delta_\mathrm{so}$. 
Because the phonon-mediated relaxation is maximal when the phonon wavelength is comparable to the size of the dots at $\omega\approx\omega_c$, an increase in $\Delta_\mathrm{so}$ leads to a decrease in $\gamma_1^\mathrm{ph}$.
Therefore, a change in $\Delta_\mathrm{so}$ as function of $\alpha$ results in the observed variations of $\gamma_\mathrm{1}^\mathrm{ph}$.
In addition, variations of $\alpha$ result in a change of the electron-phonon coupling strength that enhances this effect and is considered in $\gamma_\mathrm{1}^\mathrm{ph}(\alpha)$, plotted as black curve on top of the data in Fig.~\ref{fig:gamma}(a).

In the SM, we discuss magnetic noise. Additionally, in the SM, we discuss $1/f$-charge noise which does not capture the functional dependence of $\gamma$ as a function of the in-plane magnetic field angle $\alpha$.
\section{Conclusion}
We have investigated a singlet-triplet (S-T$^+$) qubit strongly coupled to a magnetic-field resilient microwave resonator probing the qubit.
We extract the SOI gap, the vacuum Rabi rate and the decoherence rate as function of the in-plane magnetic-field orientation.
As the central result, we find a compromise-free sweet spot at which the qubit decoherence is minimal while simultaneously the qubit transition frequency and the dipolar coupling to the resonator are maximal.
These experimental findings are well described by our theoretical model that shows that at the compromise-free sweet spot, the qubit is resilient against magnetic noise and phonon-mediated noise, with phonons forming the most likely dominant noise source.
Fitting the model, we find an anisotropic g-tensor with a maximal g-factor of $g_0=10.5$ and a spin-orbit length of $l_\mathrm{so}=120$\,nm, consistent with literature~\cite{fasth2007direct,nadj-pergeSpinOrbitQubit2010}.
Our findings demonstrate that, based on the magnetic field as global tuning parameter, optimization of speed and coherence are not mutually exclusive.
Our experimental results and model are very generic and allow for further optimization through synthesis of specifically tailored composite crystals to suppress phonon-induced loss or through a larger resonator frequency.
Sweet-spot operation might be crucial for larger qubit architectures and we anticipate that similar sweet-spots will be identified for other material platforms that rely on large spin-orbit interaction such as hole-spin qubits in Ge~\cite{hendrickx2020single,froning2021ultrafast,jirovec2021singlet,zhang2023universal}.
\section{Acknowledgements}
We are grateful for fruitful discussions with J. Danon, A. Ranni, C. Reissel and C. Ventura Meinersen.
%CS
The research was supported by the Swiss Nanoscience Institute (SNI), and the Swiss National Science Foundation through grant 192027.
This work was supported by the Swiss National Science Foundation, NCCR SPIN (grant number 51NF40-180604).
We further acknowledge funding from the European Union’s Horizon 2020 research and innovation programme, specifically the FET-open project AndQC, agreement No 828948 and the FET-open project TOPSQUAD, agreement No 847471.
We also acknowledge support through the Marie Skłodowska -Curie COFUND grant QUSTEC, grant agreement N° 847471 and the Basel Quantum Center through a Georg H. Endress fellowship.
%CT
We furthermore acknowledge support from NanoLund.
%PS
P.S. acknowledges support from the Swiss National Science Foundation through Projects No. 200021\_200418, and from the Swiss State Secretariat for Education, Research and Innovation (SERI) under contract number MB22.00081.
%DL
\section{Data availability}
The nummerical data used in this study are available in the zenodo database \url{https://doi.org/10.5281/zenodo.11205195}.
\section{Methods}
\subsection{Device fabrication}
\label{sec:methods_fab}
The fabrication process commences with the sputter deposition of approximately 10 nm of NbTiN on a pristine Si/SiO\textsubscript{2} wafer~\cite{ungerer2023performance}.
Large structures are patterned utilizing a conventional e-beam protocol and developed at room temperature. Subsequently, the narrow resonator center conductor is patterned in a second e-beam step, followed by development at $-20^\circ$C~\cite{ridderbos2018quantum,ungerer2022high}.
Following the dry-etching of the NbTiN film, a NW is deterministically deposited using a micromanipulator.
A GaSb-shell that is exculsively present on the zincblende segments of the NW~\cite{gatzke1998situ,barker2019individually}, allows us to identify the location of the wurtzite tunnel barriers using scanning electron microscopy.
After identifying the tunnel barriers, the GaSb shell is removed by wet-etching in MF-319 developer~\cite{pally2024crystal}.
Consecutively, the contacts and gates are patterned using standard e-beam lithography and thermal evaporation. For the contacts, Ar-milling in the evaporator ensures a good contact whereas the gates remain isolated by the native oxide on the NW.
\subsection{Input-output theory}
\begin{figure}
\centering\includegraphics{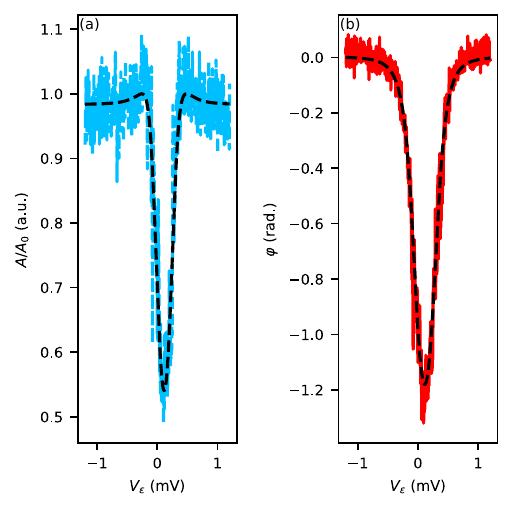}
\caption{\textbf{Linetrace with input-output theory fit at $\alpha=\pi/2$.} \textbf{(a)} Measured amplitude, $A/A_0$, as a function of the virtual gate voltage, $V_\varepsilon$, along the detuning direction. \textbf{(b)} Measured phase $\varphi$ as a function of detuning gate voltage $V_\varepsilon$.
    The data is a linecut of Fig.~\ref{fig:Brot100mT}(a) at $\alpha=\pi/2$. The dashed curves are fits to Eq.~\eqref{eq:S21}}
    \label{fig:IOfit}
\end{figure}
\label{sec:IOtheory}
To determine the SOI gap $\Delta_\mathrm{so}$, the effective spin-photon coupling strength $g_\mathrm{eff}$ and the qubit decoherence rate $\gamma$, we make use of input-output theory that describes the interaction between photons in a resonator and a multi-level system~\cite{gardiner1985input,benitoInputoutputTheorySpinphoton2017}.
Input-output theory allows us to infer the transition frequency, vacuum Rabi rate and decoherence rate of two-level systems formed in a DQD~\cite{burkard2016dispersive,mi2017high,borjans2021probing,ibberson2021large,yuStrongCouplingPhoton2023,ruckriegel2023dipole,ranni2023dephasing,ungerer2024strong}.
For the singlet-triplet (S-T+) qubit coupled to a resonator, which is probed in transmission at low power, the linear response of the measured scattering parameter is derived as~\cite{ungerer2024strong,kohler2018dispersive}
\begin{equation}
S_{21}(\omega)=\frac{-i\kappa}{\omega-\omega_0+i\frac{\kappa}{2}-g^2\left(\frac{1}{\omega-\omega_q+i\gamma}-\frac{1}{\omega+\omega_q+i\gamma}\right)} \ ,\label{eq:S21}
\end{equation}
where $\omega_q=\sqrt{\Delta_\mathrm{so}^2+{\varepsilon^\prime}^2}$, with $\varepsilon^\prime=\varepsilon-\varepsilon_0$ the relative electrostatic DQD detuning from the center of the IDT and $g=g_\mathrm{eff}\Delta_\mathrm{so}/\omega_q$.
Here we do not apply the usual rotating-wave approximation, preserving the counter-rotating terms, that are relevant for large frequency detuning $|\omega_q-\omega_0|\bcancel{\ll}|\omega_q+\omega_0|$.
To obtain the fit parameters plotted in Fig.~\ref{fig:Brot100mT}(c),~\ref{fig:Brot100mT}(d), and~\ref{fig:Brot100mT}(e) for every value of field angle $\alpha$, we fit amplitude and phase of Eq.~\eqref{eq:S21} simultaneously to the measured amplitude and phase as a function of detuning $\varepsilon$.
Fig.~\ref{fig:IOfit} shows amplitude and phase at $\alpha=\pi/2$ of the measurement shown in Fig.~\ref{fig:Brot100mT} with the overlaid fit. From this fit, we extract the SOI gap $\Delta_\mathrm{so}$, the coupling strength $g_\mathrm{eff}$, and the decoherence rate $\gamma$.
Repeating a similar procedure for various values of $\alpha$ results in the plots in Fig.~\ref{fig:Brot100mT} (c), (d) and (e). The resonator frequency $\omega_0$ and resonator decay rate $\kappa$ as well as the lever arm, transforming the applied gate voltages to detuning, are determined independently and kept fixed for the fit.
\begin{table}[tbhp]
    \centering
    \begin{tabular}{|l|}
    \hline
    Resonator\\
    \hline
    Fluctuation amplitude, $\delta\varepsilon_0 = 0.6$\\
     \hline
     Spin-orbit tensor\\
     \hline
     Spin-orbit strength, $\Theta_\mathrm{so}=d/l_\mathrm{so}=2.5$\\
     in-plane angle, $\psi=0.0$\,rad\\
     Out-of-plane angle, $\Phi=2.0$\,rad\\
     \hline
     g-tensor, $\underline{g}$\\
     \hline
         Maximal g-factor, $g_0=10.5$\\
         SOI correction, $\eta= 0.6$\\
         In-plane angle, $\psi_g=1.3$\,rad\\
         Out-of-plane angle, $\Phi_g=2.0$\,rad\\
         \hline
    Phonons\\
    \hline
    Phonon-mediated relaxation rate, $\Gamma_\mathrm{ph}/2\pi=160$\,GHz\\
    Temperature, $T=100$\,mK\\
    Normalized speed of sound, \\$c_0=c/(2\pi r)=(3\text{\,kms}^{-1})/(2\pi\cdot50\text{\,nm})$ \\
    Normalized longitudinal wavefunction extend\\ $l_0=l/r=200\text{\,nm}/50\text{\,nm}$\\
    Normalized distance between wavefunctions \\$d_0=d/r=300\text{\,nm}/50\text{\,nm}$\\
    \hline
    \end{tabular}
    \caption{Free parameters used to fit the data shown in Figure 3 of main text. The tunnel rate $t_c/2\pi=4.5$\,GHz as well as the gate-to-dot lever arms $\alpha_{L2}= 0.014\pm0.003$, $\alpha_{L1}=0.044\pm0.008$, $\alpha_{R2}= 0.29\pm 0.06$ and $\alpha_{R2}= 0.045\pm0.014$ were measured independently, where $\alpha_{ij}$ is the dimensionless lever arm between gate $i$ and dot $j$.
    The spin-orbit strength $\Theta_\mathrm{so}$ combined with the distance between the dots $d=300$\,nm (center to center) corresponds to a spin-orbit length $l_\mathrm{so}=d/\Theta_\mathrm{so}\approx120$\,nm. The last three fit parameters are normalized by the wavefunction radius, $r=50$\,nm.}
    \label{tab:params}
\end{table}
\begin{figure}
\centering
\includegraphics[]{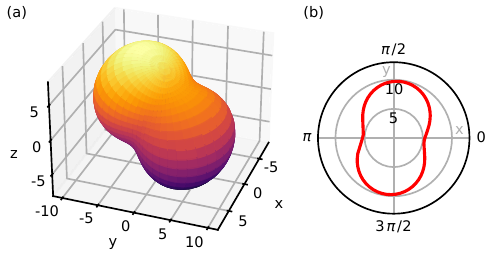}
\caption{\textbf{Visualization of the anisotropic g-tensor.} \textbf{(a)} Normalized Zeeman field strength $|\mathbf{B}\underline{g}|/|\mathbf{B}|$, where the g-tensor $\underline{g}$ is defined according to Eq.~\eqref{eq:gtensor}. A characteristic peanut shape is exhibited. \textbf{(b)} Cross-section of (a) in the $x$-$y$ plane at $z=0$ as a function of the in-plane angle $\alpha$. The nanowire is aligned parallel to the $y$ axis.}
    \label{fig:g-tensor}
\end{figure}
\subsection{Model of the double quantum dot}
The curves plotted in Fig.~\ref{fig:Brot100mT} and~\ref{fig:gamma} of the main text are the result of numerical diagonalization of the effective Hamiltonian of a double quantum dot (DQD) with even occupation close to the avoided crossing between $|T_{1,1}^{+}\rangle$ and $|S_{2,0}\rangle$.
The $(1,1)\to (0,2)$ anticrossing is driven by changing the detuning $\varepsilon$ between the two dots.
For the usual negative exchange splitting between S and T states, the (1,1) state is preferred at $\varepsilon<0$ and the state (2,0) is preferred at $\varepsilon > 0$.
 
The singlet and triplet states in the DQD are $(| S_{0,2}\rangle,| S_{2,0}\rangle,| S_{1,1}\rangle,| T_{1,1}^{-}\rangle,| T_{1,1}^{+}\rangle,| T_{1,1}^{0}\rangle)$ and form our basis set.
The Hamiltonian including SOI then reads~\cite{stepanenko2012singlet,spethmann2024high,geyer2024anisotropic}
\begin{widetext}
\begin{equation}
H_\text{DQD} = 
\left(
\begin{array}{cccccc}
 U_{L}+U_R +\varepsilon & U_{LR} & -t_c  & 0 & 0 & 0 \\
 U_{LR} & -\varepsilon & -t_c  & 0 & 0 & 0 \\
- t_c  &  -t_c & -J_0 & -\frac{\delta b_x+i\delta b_y}{\sqrt{2}} & \frac{\delta b_x-i\delta b_y}{\sqrt{2}} & \delta b_z \\
0 & 0 & -\frac{\delta b_x-i\delta b_y}{\sqrt{2}} & \bar{b}_z & 0 & \frac{\bar{b}_x-i\bar{b}_y}{\sqrt{2}} \\
 0 & 0 & \frac{\delta b_x+i\delta b_y}{\sqrt{2}} & 0 & -\bar{b}_z & \frac{\bar{b}_x+i\bar{b}_y}{\sqrt{2}} \\
 0 & 0 & \delta b_z & \frac{\bar{b}_x+i\bar{b}_y}{\sqrt{2}} & \frac{\bar{b}_x-i\bar{b}_y}{\sqrt{2}} & 0 \\
\end{array}
\right) \ .
\label{eq:H6x6}
\end{equation}
\end{widetext}
Here, $U_{L/R}$ is the on-site Coulomb energy of the left (L) and right (R) dot, $U_{LR}$ is the cross-exchange contribution, and $J_0$ is the Coulomb exchange.
And we define the difference and total Zeeman field vectors produced by a magnetic field $\textbf{B}$ as
\begin{align}
\label{eq:deltab}
\delta\textbf{b}&={\mu_B} \textbf{B} \delta{g} =\frac{\mu_B}{2} \textbf{B} \Big[\underline{g}_L\underline{R}(-\theta_{so}/2)-\underline{g}_R\underline{R}(\theta_{so}/2) \Big] \ , \\
\label{eq:bbar}
  \bar{\textbf{b}}&={\mu_B} \textbf{B} \bar{g}=\frac{\mu_B}{2} \textbf{B} \Big[\underline{g}_L\underline{R}(-\theta_{so}/2)+\underline{g}_R\underline{R}(\theta_{so}/2) \Big]\ ,
\end{align}
where $\underline{R}$ is a rotation matrix around the SOI axis by the spin-flip angle $\theta_{so}=d/l_{so}$, with $d$ the distance between the dots and $l_{so}$ the SOI length.
For simplicity, we set $U_{LR}=J_0=0$.
Following Ref.~\cite{geyer2024anisotropic}, we have gauged the spin-flip tunneling contribution into a local redefinition of the Zeeman energy and thus the tunneling amplitude $t_c$ comprises both spin-conserving and spin-flip components.
$\underline{g}_{L/R}$ are the g-tensors of the left and right dots, parameterized by the unit vector $\mathbf{n}$ and the SOI correction factor $\eta\leq1$,
\begin{equation}
    \mathbf{B}\underline{g} = g_0 
    \left[\left(\mathbf{n}(\mathbf{n}\cdot\mathbf{B})\right)+\eta(\mathbf{n}\times\mathbf{B})\times \mathbf{n}\right] \ .
    \label{eq:gtensor}
\end{equation}
For describing the experimental results and to limit the number of free parameters, we use identical g-tensors for the left and right dot, $\underline{g}=\underline{g}_{L}=\underline{g}_{R}$.

The DQD is coupled to a resonator of frequency $\omega_0$ and bosonic annihilation and creation operators $a$ and $a^\dagger$, with the effective Hamiltonian $ H_R=\hbar\omega_0 a^\dagger a $.
The coupling is mediated by the resonator-induced change in detuning of the DQD and thus reads
\begin{align}
\label{eq:coupling}
  H_I&= \delta\varepsilon (|S_{20}\rangle\langle S_{20}|-|S_{02}\rangle\langle S_{02}|)  (a^\dagger+a) \ ,
  \end{align}
where $\delta\varepsilon=V_R \partial_V\varepsilon $ depends on the susceptibility of the detuning to the potential applied to the gate and on the zero-point-fluctuations $V_R$ of the potential of the resonator.

Focusing on the $(1,1)\to (0,2)$ anticrossing at $\varepsilon\sim \varepsilon_0$, we  neglect the contribution of the singlet $|S_{20}\rangle$ which lies at high energy $U_R+U_L$.
We can then easily diagonalize the singlet subsector by accounting for the tunneling between $S_{11}$ and $S_{20}$, and we focus on the 5 lowest energy states of the system, comprising the resulting singlets and the three triplets.
The effective Hamiltonian reads  
\begin{widetext}
\begin{equation}
\label{eq:H5}
H_5=\left(
\begin{array}{ccccc}
J-\varepsilon& 0 & -\frac{(\delta b_x'+i \delta b_y') \cos (\theta )}{\sqrt{2}} & \frac{(\delta b_x'-i \delta b_y') \cos (\theta )}{\sqrt{2}} & \delta b_z' \cos (\theta ) \\
 0 & -J & -\frac{(\delta b_x'+i \delta b_y') \sin (\theta )}{\sqrt{2}} & \frac{(\delta b_x'-i \delta b_y') \sin (\theta )}{\sqrt{2}} & \delta b_z' \sin (\theta ) \\
 -\frac{(\delta b_x'-i \delta b_y') \cos (\theta )}{\sqrt{2}} & -\frac{(\delta b_x'-i \delta b_y') \sin (\theta )}{\sqrt{2}} & E_z & 0 & 0 \\
 \frac{(\delta b_x'+i \delta b_y') \cos (\theta )}{\sqrt{2}} & \frac{(\delta b_x'+i \delta b_y') \sin (\theta )}{\sqrt{2}} & 0 & -E_z &0 \\
 \delta b_z' \cos (\theta ) & \delta b_z' \sin (\theta ) & 0& 0 & 0 \\
\end{array}
\right) \ ,
\end{equation}
\end{widetext}
\normalsize
where $J\approx (\varepsilon+\sqrt{\varepsilon^2+4t_c^2})/{2}$ according to Eq.~\eqref{eq:EzeqJ} and $\theta=\text{arctan}(2t_c/\varepsilon)/2$, where the arctan function is defined, such that $\theta=\pi/4$ at $\varepsilon=0$.
We also fixed the direction of the spin quantization axis such that the triplet subsector is diagonal.
This is done by defining the global rotation matrix $\underline{R}_B$ that maps $\bar{\textbf{b}}$ to the $z$-direction and rotating the $\delta \textbf{b}$ accordingly, i.e.
\begin{equation}
\bar{\textbf{b}}=E_z \underline{R}_B \textbf{n}_z \ \to \ \ \delta{\textbf{b}}'=\delta{\textbf{b}}\underline{R}_B \ .
\end{equation}
Here, $\underline{R}_B$  depends on the direction of the applied $\textbf{B}$ field and on the combined g-tensor of the dots.
The prime in $\delta \textbf{b}$ indicates this choice of reference frame for the spin quantization axis. 

The resonator-DQD coupling Hamiltonian $H_I$ in Eq.~\eqref{eq:coupling} is related to variations in the detuning $\delta\epsilon$ by
\begin{equation}
  H_I=  \delta\varepsilon_0 h_\varepsilon
(a^\dagger+a),
\end{equation}
with 
\begin{equation}
    h_\varepsilon=\left(
\begin{array}{ccccc}
 -\sin ^2(\theta ) & \sin (2\theta )/2 & 0 & 0 & 0 \\
 \sin (2\theta )/2 & -\cos ^2(\theta ) & 0 & 0 & 0 \\
 0 & 0 & 0 & 0 & 0 \\
 0 & 0 & 0 & 0 & 0 \\
 0 & 0 & 0 & 0 & 0 \\
\end{array}
\right) \ 
\label{eq:delta_det}
\end{equation}
describing small variations of the detuning $\varepsilon$. This leads to the definition of the dipolar coupling strength according to Eq.~\eqref{eq:geff}.
\bibliographystyle{apsrev4-1} % Tell bibtex which bibliography style to use
\bibliography{bib}
\end{document}